\documentstyle[prl,aps,epsbox,multicol]{revtex}

\newcommand{\eq}[1]{(\ref{#1})}

\begin{document}
\draft
\preprint{draft}
\twocolumn[\hsize\textwidth\columnwidth\hsize\csname@twocolumnfalse\endcsname

\title{Aging Effects of an Elastic String
Diffusing in a Disordered Media}

\author{Hajime Yoshino}

\address{Institute for Solid State Physics, the Univ. of Tokyo,
      7-22-1 Roppongi, Minato-ku, Tokyo 106-8666 Japan}

\maketitle

\begin{abstract}
The aging effects of a 'diffusing glass', a single elastic string 
diffusing in a two-dimensional disordered media 
assisted by thermal noise are studied by Monte Carlo (MC) simulations.
We find for the first time convincing numerical evidence 
of non-trivial aging effects  both in the 
linear response against transverse force and the associated
correlation function. 
Our results retain some important predictions of the conventional 
dynamical mean-field theory (MFT) but also 
reveal remarkable differences from it.  

\end{abstract}


\vskip1pc]
\narrowtext
Elastic objects diffusing in disordered media assisted by
thermal noise, which we call simply as 'diffusing glasses', 
are expected to remain persistently {\it non-stationary}
due to the complicated competitions between the elasticity and
disorder. An example will be a single vortex line 
diffusing in a type-II super-conductor with 
randomly distributed pinning centers.
A dynamical MFT of elastic 
manifolds in random media \cite{CKD96} has been developed which 
proposes a self-consistent picture for such dynamics and
predicts existence of aging effects similar to those found 
in spin-glasses.\cite{sgexp} However, the MFT is exact only
in infinite dimensional embedding space. 
Compared with many {\it driven} dynamics \cite{blatt} 
of the same elastic objects, much less is known about 
such diffusive dynamics in realistic dimensional spaces.

In order to go beyond the MFT,
we study in this letter the dynamics of a  simplest 'diffusing glass',
namely directed polymer in random media (DPRM) \cite{HZ95} 
on a two dimensional square lattice by extensive MC simulations. 
The previous MC simulations \cite{Y96,B97} 
have found encouraging signatures of non-stationary dynamics 
under zero driving force but many open questions remained.

We find that the $k=0$ component of current (velocity) correlation function 
and linear conductivity exhibit novel scaling behaviors 
which reveal 'one step' violations of
the time translational invariance (TTI) 
and the Green-Kubo (GK) formula or
the fluctuation dissipation theorem (FDT) for 'currents'\cite{KTH}.
On the other hand, finite $k$ components equilibrate in the time 
 $\tau_{k} \propto k^{-z(T)}$.

{\em Model}.--
Our model is a lattice string on a two dimensional 
square lattice of size $L$ in the longitudinal  
direction $z$ and $M$ in the transverse direction $x$.
The string is directed in $z$ direction and its segments
are labeled as $x(z)$ ($z=1,2,\cdots,L$). They obey the RSOS 
(Restricted Solids on Solid) condition, i.e. $|x(z)-x(z-1)|=0,\pm1$.
Random potential $V(z,x)$ is defined on each lattice site $(z,x)$ 
which takes a random value from the uniform distribution
between -1 and 1. This system is known to be
in the glassy phase  at all finite temperatures.\cite{HZ95}

The dynamics is introduced by a heat-bath type MC method \cite{Y96,B97}
which ensures that the microscopic motions of 
the {\it kinks} and {\it anti-kinks}
are thermalized with the heat-bath at a  temperature $T_{\rm bath}$.
In one Monte Carlo Step (MCS), the whole
configuration is sweeped once. 
The energy of the system at time $t$ is given by
\begin{equation}
E(t)[V,h] =  \sum_{z=1}^{L} V(z,x(z,t))-h(z,t)x(z,t),
\end{equation}
where $h(z,t)$ is the transverse force.
As in the case of static properties \cite{HZ95}, we assume that 
the dynamics of the present lattice model and the 
continuous model studied in the MFT have  same
asymptotic scaling properties.

We have performed simulations similar to the zero field cooling (ZFC) 
experiments of spin-glass systems. \cite{sgexp}. At first, 
a certain {\it waiting time}  $t_{2}$ is elapsed under zero field
starting from an out-of equilibrium initial configuration. 
Then a constant filed $h(z')$ is applied at $z'$ afterwards.

If linear response holds, the induced {\it current} at $z$, i. e. 
temporal transverse velocity of the segment $x(z)$, to be measured 
at time $t_{1}( > t_{2})$  can be written as,
\begin{equation}
\delta J_{z-z'}(t_{1},t_{2})
=\int_{t_{2}}^{t_{1}}dt' \sigma_{z- z'}(t_{1},t')h(z'),
\end{equation}
where $\sigma_{z- z'}(t,t')$ ($t > t'$) is a time-dependent conductivity.
Instead of the currents, we measure the linear
susceptibility (induced displacement divided by $h$), 
\begin{eqnarray}
\chi_{z-z'}(t_{1},t_{2}) & = &
\int_{t_{2}}^{t_{1}} dt  \frac{\delta J _{z-z'}(t,t_{2})}{h(z')}
\nonumber \\
& = &  \int_{t_{2}}^{t_{1}}dt \int_{t_{2}}^{t}dt' \sigma_{z- z'}(t,t'). \label{chi}
\end{eqnarray}

Let us here introduce a generalized Green-Kubo (GK) formula, 
\begin{equation}
\sigma_{z-z'}(t_{1},t_{2}) = 
\frac{Y_{z-z'}(t_{1},t_{2})}{T_{\rm bath}}<J_{z}(t_{1})J_{z'}(t_{2})>
\theta(t_{1}-t_{2}), 
\label{kubo1}
\end{equation}
where
$<J_{z}(t_{1})J_{z'}(t_{2})> \equiv \partial_{t_{1}}\partial_{t_{2}}
<x(z,t_{1})x(z',t_{2})>$ 
is the current correlation function. The usual GK formula 
\cite{KTH} corresponds to the case with the 'FDT ratio' $Y=1$.
Hereafter, the bracket $<\cdots>$ means the average over 
samples: different realizations of initial configurations, thermal histories
(MC runs) and random potentials.

By integrating the current correlation function
over the two time variables we obtain,
\begin{eqnarray}
B_{z-z'}(t_{1},t_{2}) & \equiv & 
\int_{t_{2}}^{t_{1}}dt \int_{t_{2}}^{t_{1}} dt'
<J_{z}(t)J_{z'}(t')> \nonumber \\
&=& <[x(z,t_{1})-x(z,t_{2}) ]
[x(z',t_{1})-x_(z',t_{2})]>.
\end{eqnarray}
Combining with \eq{chi}, the {\it integral} violation 
of the FDT can be defined as, 
\begin{equation}
I_{z-z'}(t_{1},t_{2}) 
= B_{z-z'}(t_{1},t_{2})/2-T_{\rm bath}\chi_{z-z'}(t_{1},t_{2}).
\label{kubo2}
\end{equation}

We define the Fourier transform along $z$ axis as
$\chi_{k}(t_{1},t_{2}) =
\sum_{r=0}^{L}\cos(kr)\chi_{r}(t_{1},t_{2})$
and
$B_{k}(t_{1},t_{2}) = \sum_{r=0}^{L}\cos(kr)B_{r}(t_{1},t_{2})$
where $k=n \pi /L$ with ($n=0,1,\cdots,L$).
Note that $B_{k=0}$ is proportional to the mean-squared displacement
of the center of mass.

In the following, we begin with the crossover between
the linear and non-linear response.
Then, we discuss aging effects in linear responses and
correlation functions.

{\em Crossover between linear and  non-linear response}.--
As often emphasized, the non-linear 
effects due to the collective creep
\cite{blatt} should be dominant asymptotically. However
one can systematically expel them out of a given time 
window by decreasing $h$  because the characteristic time to create
the nucleus of the creep diverges rapidly  as $h$ is decreased by 
the well known formula $\ln \tau_{\rm creep} \propto h^{-\mu}$ 
with the glassy exponent $\mu=1/4$  in two dimension.
We demonstrate it in the following.

We have performed simulations of system size $L=500$ $M=520$, at 
temperature $T_{\rm bath}=0.4$ under {\it uniform} transverse fields.
The periodic boundary condition 
is imposed on both $x$ and $z$ directions and 
straight lines are chosen as initial configurations.

In Fig.~1, we show data of the susceptibility $\chi_{k=0}(\tau,0)$  
at different strength of fields $h$ in a double logarithmic
plot. There are crossovers from the linear response curve
where data of different $h$ merge with each other
to the non-linear branches which depart from the common curve.  
The non-linear branch grow linearly with $\tau$ by
different velocities $v(h)$.
In the inset of  Fig.~1 we show the effective velocity $v(h)$ 
determined by fitting the non-linear branches to the form  
$h\chi(\tau,0)=v(h)\tau+c(h)$ where $c(h)$ is a parameter.
The result  is consistent with the expected behavior
$\ln v(h) \propto - \ln \tau_{\rm creep}$. More details of the
crossover phenomena will be reported elsewhere. 

{\em Aging Effects}.-
Let us begin discussions on aging effects
in the linear susceptivility $\chi$ and integrated
correlation function $B$ at zero wavenumber $k=0$.

In Fig.~2  we show the data of 
$\chi_{k=0}$ and $B_{k=0}$
of different waiting times $t_{2}$
against the time difference $\tau=t_{1}-t_{2}$
in double logarithmic plots. 
The system size is $L=500$ and temperature is $T_{\rm bath}=0.4$.
We have checked that 
there are no finite size effects up to $10^{6}$ MCS 
within the statistical accuracy by simulating larger systems.
We have chosen weak enough uniform field  $h=0.005$ in order to
measure linear responses up to $10^{6}$ MCS.

\begin{figure}[t]
\hspace*{0.2cm}
\psbox[width=70mm,aspect=1.2]{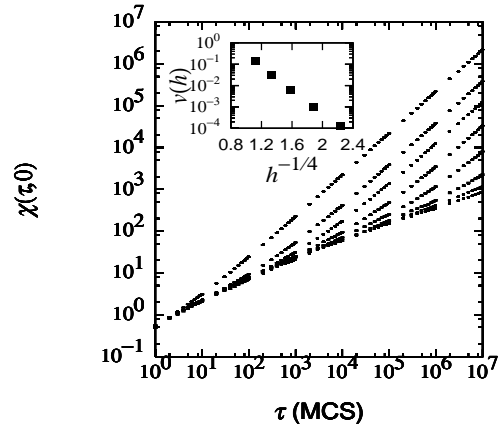}
\label{crossover.fig}
\caption{Crossover between linear and non-linear responses.
The strength of the uniform field is varied as
$h=0.64$ ,$0.32$, $0.16$, $0.08$, $0.04$, $0.02$, 
$0.01$, $0.005$, $0.0025$ from the top curve to the bottom. 
The error bars are of the size of the symbols.
The average is took over $100$ samples. 
($1000$ samples for $h=0.005$ and $0.0025$)
The inset is the velocity $v(h)$ versus  $h^{-1/4}$
at  $h=0.64$ ,$0.32$, $0.16$, $0.08$, $0.04$.
}
\end{figure}

Except the rapid growth at $\tau <10^{2}$ MCS,  which 
is discarded in the following as a short time transient behavior,
the general feature is the following: 
each curve of a given waiting time $t_{2}$ follows
initially the lower 'quasi-equilibrium' branch 
then switches over to the upper 'aging' branch at around 
$\log \tau \sim \log t_{2}$.  
Here the existence of aging effects is evident. 
Both $\chi$ and $B$ violates TTI:  they are not functions 
of the time difference $\tau\equiv t_{1}-t_{2}$ alone but 
explicitly depends also on the {\it waiting time} $t_{2}$.

\begin{figure}[t]
\psbox[width=85mm,aspect=1.6]{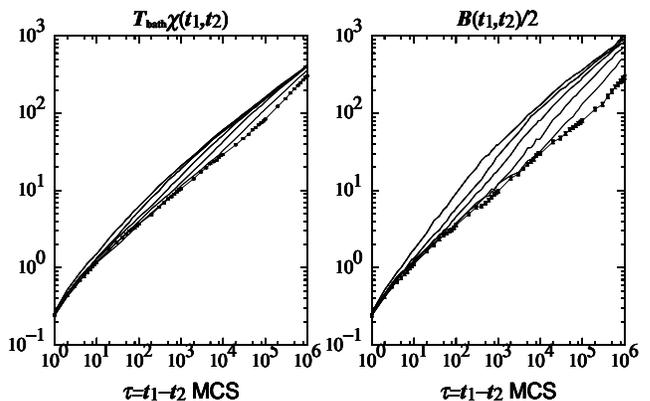}
\label{chi.fig}
\caption{$T_{\rm bath}\chi_{k=0}(t_{1},t_{2})$ and
$B_{k=0}(t_{1},t_{2})/2$ of 
waiting times $t_{2}=10$, $10^{2}$, $10^{3}$, $10^{4}$, $10^{5}$, $10^{6}$
from the left curve to the right. 
Only lines connecting the data points 
at $\tau=t_{1}-t_{2}=n\times 10^{p}$ 
(where $n=1,2,\cdots,9$ and $p=0,1,\cdots,5$)
are shown for graphical convenience. 
The average is took over 1050 samples.
Typical sizes of the error bars are those of the symbols 
on the curve of $t_{2}=10^{6}$ (bottom).}
\end{figure}

\begin{figure}[t]
\hspace*{5mm}
\psbox[width=65mm,aspect=1.0]{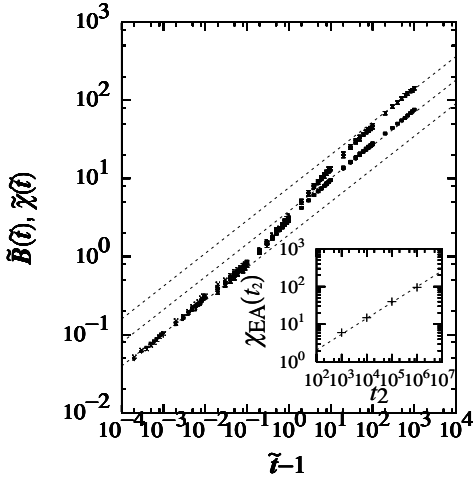}
\label{scale.fig}
\caption{Scaling plot of the susceptibility and the mean-squared
displacement. (lower and upper curves respectively)
 The inset is the scaling parameter
$\chi_{\rm EA}(t_{2})$ vs $t_{2}$. 
We use  the data of $\tau > 10^{2}$ (MCS) at
$t_{2}=10^{3}$,$10^{4}$,$10^{5}$,$10^{6}$ (MCS).
All the broken lines are 
algebraic law fits of the same exponent $2/z(T_{\rm bath}=0.4)=0.42$ with 
different amplitudes, from which we obtain 
$c_{2}/c_{1}\simeq 4.0$ and $y\simeq 0.5$.}
\end{figure}

\begin{figure}[t]
\hspace*{1cm}
\psbox[width=65mm,aspect=1.0]{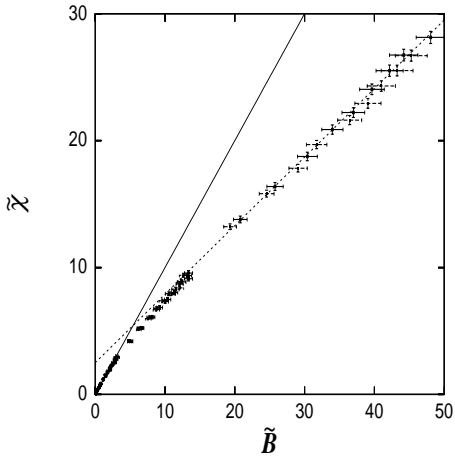}
\label{parametric.fig}
\caption{Parametric plot of the scaling functions 
$\tilde{\chi}(\tilde{t})$ and $\tilde{B}(\tilde{t})$ shown in Fig.~3.
The solid straight line represents the integrated FDT. 
The dashed line is a fit by a straight line with slope $y=0.54$
to the part where the FDT is violated.}
\end{figure}

The {\it one step} structures
can be well described by the following simple scaling ansatz:
\begin{eqnarray}
T_{\rm bath}\chi_{k=0}(t_{1},t_{2}) & \simeq &
S_{\chi}(t_{1}/t_{2}) (t_{1}-t_{2})^{2/z(T)}\nonumber \\
B_{k=0}(t_{1},t_{2})/2 
&\simeq&S_{B}(t_{1}/t_{2}) (t_{1}-t_{2})^{2/z(T)}
\label{scaletime}.
\end{eqnarray}
Here $S_{\chi}(\tilde{t})$, $S_{B}(\tilde{t})$
are step-like functions in the scaled time variable
$\tilde{t}=t_{1}/t_{2}$. They take a) the same value
$c_{1}$ in the 'quasi-equilibrium time domain' $\tilde{t} \ll 1$ and
b) different values $y c_{2}( > c_{1})$ and $c_{2}(> c_{1})$ 
respectively in the 'aging time domain' $\tilde{t} \gg 1$. 
The ratio $y$ is smaller than $1$. The latter means that
the integral violation of FDT \eq{kubo2} exhists : 
$I > 0$ in the 'aging time domain'.

In the absence of the random potential, the present model is
equivalent to the free Gaussian field model \cite{CKP94}
in which  $z=2$ : the diffusion is 'normal'.
Further more $S_{\rm \chi}$ and $S_{\rm B}$ are constants
(mobility or diffusion constant) and equal. Thus TTI and the FDT 
holds completely.

In order to test the scaling ansatz \eq{scaletime}, 
we consider equivalent formulas: 
$T_{\rm bath}\chi_{k=0}(t_{1},t_{2})\simeq \chi_{\rm EA}(t_{2})\tilde{\chi}(\tilde{t})$ 
and $B_{k=0}(t_{1},t_{2})/2 \simeq \chi_{\rm EA}(t_{2})\tilde{B}(\tilde{t})$ 
where  $\chi_{\rm EA}(t_{2}) = t_{2}^{2/z(T)}$ scales the 'height of
step'. The scaling functions read as
$\tilde{\chi}(\tilde{t})=S_{\chi}(\tilde{t})|\tilde{t}-1|^{2/z(T)}$
and $\tilde{B}(\tilde{t})=S_{B}(\tilde{t})|\tilde{t}-1|^{2/z(T)}$.
We assume that $\tilde{t}=t_{1}/t_{2}$ is the correct scaling variable and
treat $\chi_{\rm EA}(t_{2})$ as a scaling parameter \cite{CK}
to be determined for each $t_{2}$.
The resultant master curves and the scaling parameters 
$\chi_{\rm  EA}(t_{2})$ are shown in Fig.~3.
In Fig.~4, we show a parametric plot of the master curves which 
clearly shows 'one step' violation of the FDT.

From \eq{scaletime}, we can deduce scaling forms for the 
linear conductivity and correlation function
of the current at $k=0$.
Except around the {\it step} $\log (t_{1}/t_{2}-1) \sim 0$, we 
find power laws multiplied by the 'one step' functions,
\begin{eqnarray}
T_{\rm bath}\sigma(t_{1},t_{2}) & \simeq &
S_{\chi}(t_{1}/t_{2})(t_{1}-t_{2})^{-2(1-1/z(T))} \nonumber \\
<J(t_{1})J(t_{2})>  & \simeq & 
S_{B}(t_{1}/t_{2})(t_{1}-t_{2})^{-2(1-1/z(T))}. \label{kubo3}
\end{eqnarray}
In the last equations we have used 
$\sigma(t_{1},t_{2})=-\partial_{t_{1}}\partial_{t_{2}}\chi(t_{1},t_{2})$,
$<J(t_{1})J(t_{2})>=-\partial_{t_{1}}\partial_{t_{2}}B(t_{1},t_{2})/2$ and 
omitted the common pre-factor $-2/z(T)(1-2/z(T))$.
So both {\it noise} and {\it response} violate TTI and the FDT.
The 'FDT ratio'  $Y(t_{1},t_{2})$ \eq{kubo1} shows 'one step'
variation : $Y=1$ in the the 'quasi-equilibrium time domain'
and $Y=y$ in the 'aging time domain'.

Let us now compare our results with the conventional picture based
on the MFT \cite{CKD96}. 
There are now increasing number of numerical studies 
on different glassy systems 
such as a spin-glass model\cite{FR95}, 
Lennard-Jones Glasses\cite{P97}, and coarsening in spin-systems\cite{B98},
which appear to {\it roughly} support the picture based on 
similar dynamical MFTs \cite{BCKM}. 
However we find that the crucial concept of the MFTs \cite{CKD96,BCKM}
called as 'correlation scales'
cannot be applied to our system as we discuss below. 

The MFT assumes that 
TTI and the FDT hold in the quasi-equilibrium scale $B < B_{\rm EA}$
but not in the aging scale $B > B_{\rm EA}$.
The border line between the two scales $B_{\rm EA}$ is 
a well-defined constant.
Consequently, the correlation and response functions 
becomes the {\it sum} of the contributions from the 
quasi-equilibrium and aging scale.  

However, the {\it multiplicative} scaling form \eq{scaletime} 
certainly disagree with the MFT.
The 'height of step' $\chi_{\rm  EA}(t_{2})$, 
which corresponds to the
$B_{\rm EA}$ of the MFT, is {\it not} at all a constant 
but apparently increases with time $t_{2}$ as shown in the inset of Fig. 3
(similar effect has been noticed in the Sinai model \cite{LD97}).
Thus our results reveal considerable differences from the MFT.
However, surprisingly, the 'one step' variation of the
FDT ratio predicted by the MFT is recovered in an unexpected way.

{\em Propagation of response}.--
Finally we discuss the propagation of responses along $z$ axis.
To this end, we have performed another set of simulations 
with {\it point} fields $h(z)=h_{\rm point}\delta(z-L/2)$
which pull the center of the string $z=L/2$.
We have obtained $\chi_{r}(t_{1},t_{2})$
by measuring the induced displacement of segments 
at various distance $r$ from the center.

In Fig.~5. we show a set of data  in a scaling plot
in order to test the  scaling law
which holds in the pure system \cite{CKP94},
\begin{equation}
\chi_{r}(t_{1},t_{2}) =
\chi_{r=0}(t_{1},t_{2})H(r/\chi_{r=0}(t_{1},t_{2})).
\label{spacescale}
\end{equation}
Here $\chi_{r=0}(t_{1},t_{2})$ plays the role very similar to 
the 'domain size' in the {\it coarsening } systems \cite{B94}.
In the scaling analysis, we have treated each $\chi_{r=0}(t_{1},t_{2})$ 
as a scaling parameter to be determined for each $(t_{1},t_{2})$.

\begin{figure}[b]
\hspace*{8mm}
\psbox[width=65mm,aspect=1.3]{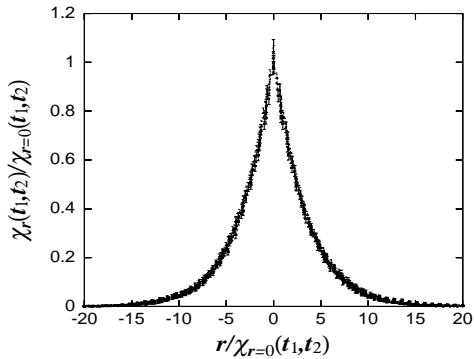}
\label{space.fig}
\caption{Scaling plot of the profile
$\chi_{r}(t_{1},t_{2})$ at 
$t_{2}=10^{3}$, $10^{4}$, $10^{5}$, $10^{6}$ 
and $\tau=10^{3}$, $10^{4}$, $10^{5}$, $10^{6}$.
The system size is  $L=500$ and  temperature is $T_{\rm bath}=0.4$
The average is took over $11520$  samples.
The non-linear effect is much weaker than in the case of uniform fields and 
we have used safely point field of strength $h_{\rm point}=0.1$ 
to investigate linear-response up to $10^{6}$ MCS. 
}
\end{figure}

The resultant master curve $H(\tilde{r})$ drops linearly
$H(\tilde{r})= 1-s \tilde{r}$ 
at $\tilde{r} < 2$ and has a Gaussian tail
at $\tilde{r} > 10$.  Surprisingly, it is 
virtually indistinguishable from that obtained
by simulating a corresponding pure system within our numerical
accuracies.

In the Fourier space, the scaling form \eq{spacescale} takes
the same form as in the MFT\cite{CKD96},
$k^{2}\chi_{k}(t_{1},t_{2}) = F(k^{2}\chi_{k=0}(t_{1},t_{2}))$
where $\chi_{k=0}(t_{1},t_{2})=h\chi_{r=0}^{2}(t_{1},t_{2})$.
(Our data satisfy latter within our numerical accuracies.)
In the last equations we used $h\equiv\int_{0}^{\infty}dyH(y)$ 
and $F(x)\equiv (x/h)\int_{0}^{\infty} dy \cos(y\sqrt{x/h})H(y)$.
Combining with \eq{scaletime} we find that 'domain size' grows as
$\chi_{r=0}(t_{1},t_{2})) \simeq \sqrt{S_{\chi}(t_{1}/t_{2})/h}
(t_{1}-t_{2})^{1/z(T)}$. 
In other words, a finite $k$ component equilibrates
in the time  $\tau_{k} \propto k^{-z(T)}$
to the equilibrium value $\chi_{k}^{\rm eq} \propto k^{-2}$.
The latter $k^{-2}$ scaling is due to the 
so called statistical tilt symmetry \cite{HF94}.

The exponent $z(T_{\rm bath}=0.4)\simeq 4.8$ 
is much larger than in the  pure model $z=2$.
We have found that $z(T)$
increases with decreasing temperature. 
Though the previous work \cite{Y96} proposed a
logarithmic law for the susceptibility
the present data of increased statistical accuracy 
over enlarged time range fit  better to the algebraic law
in \eq{scaletime}. Thus the naive scaling argument based only on 
{\it typical} energy barrier \cite{Y96} which suggest
the logarithmic law should be inaccurate.

To summarize, we performed extensive MC simulations
and scaling analysis on the aging effects of a 
'diffusing glass', 2 dimensional
DPRM. We found that the correlation and linear response functions 
exhibit  interesting 'one-step' scaling phenomena.
It will be very interesting if such features can be observed 
experimentally in some 'diffusing glasses', such as vortex lines 
in dirty type-II super-conductors.

The author gratefully thanks A. Barrat, J. P. Bouchaud, 
L. F. Cugliandolo,
M. Hamman, J. Kurchan,  H. Rieger and H. Takayama  
for discussions and suggestions.
This work was supported by Grand-in-Aid for Scientific Research 
from the Ministry of Education, Science and Culture, Japan.
The computation has been done using the facilities 
of the Supercomputer Center, ISSP,
the Univ. of Tokyo, the Computer Center, the Univ. of Tokyo
and the Computer Center, Kyushu University.

\end{document}